\documentclass[manuscript]{aastex62}

\usepackage{graphicx}
\usepackage{booktabs}
\usepackage{scalefnt}

\newcommand{\SAO}{Universidade de S\~ao Paulo, IAG, Rua do Mat\~ao 1226, Cidade Universit\'aria, S\~ao Paulo 05508-900, Brazil}    

\newcommand{\UBO}{Centro de Investigaci\'on en Astronom\'ia, Universidad Bernardo O’Higgins, Avenida Viel 1497, Santiago, Chile}
\usepackage{comment}

\begin{document}

\title{Phosphorous in the moderately metal-poor bulge globular clusters NGC~6539 and NGC~6569}

\correspondingauthor{Beatriz Barbuy}
\email{b.barbuy@iag.usp.br}

\author[0009-0002-4599-7185]{Morgan S. Camargo}
\affil{\SAO}

\author[0000-0001-9264-4417]{Beatriz Barbuy}
\affil{\SAO}

\author[0000-0001-6541-1933]{Heitor Ernandes}
\affil{Nicolaus Copernicus Astronomical Center, Polish Academy of Sciences, ul. Bartycka 18, 00-716 Warsaw, Poland}

\author[0009-0007-8204-1234]{Am\^ancio C. S. Fria\c ca}
\affil{\SAO}

\author[0000-0003-1269-7282]{Cristina Chiappini}
\affil{Astrophysikalisches Institut Potsdam, An der Sternwarte 16, Potsdam, 14482, Germany}

\author[0000-0002-7552-3063]{Bernardo P. L. Ferreira}
\affil{\SAO}

\author[0000-0001-8052-969X]{Stefano O. Souza}
\affil{Max Planck Institute for Astronomy, K\"onigstuhl 17, D-69117 Heidelberg, Germany}


\author[0000-0001-7939-5348]{Sergio Ortolani}
\affil{Universit\`a di Padova, Dipartimento di  Fisica e Astronomia, Vicolo dell'Osservatorio 2, I-35122 Padova, Italy}
\affil{Centro di Ateneo di Studi e Attivit\`a Spaziali “Giuseppe Colombo” – CISAS, Via Venezia 15, 35131 Padova, Italy}
\affil{INAF-Osservatorio di Padova, Vicolo dell'Osservatorio 5, I-35122 Padova, Italy}

\author[0000-0003-1149-3659]{Domenico Nardiello}
\affil{Universit\`a di Padova, Dipartimento di  Fisica e Astronomia, Vicolo dell'Osservatorio 2, I-35122 Padova, Italy}
\affil{Centro di Ateneo di Studi e Attivit\`a Spaziali “Giuseppe Colombo” – CISAS, Via Venezia 15, 35131 Padova, Italy}

\author[0000-0003-3336-0910]{Eduardo Bica}
\affil{Universidade Federal do Rio Grande do Sul, Departamento de Astronomia, CP 15051, Porto Alegre 91501-970, Brazil}

\author[0000-0003-3526-5052]{Jos\'e G. Fern\'andez-Trincado}
\affil{\UBO}

\author[0009-0008-1489-972X]{Nicolas Barrera}
\affil{Departamento de Astronomía, Facultad de Ciencias, Universidad de La Serena, Av. Juan Cisternas 1200, La Serena, Chile}

\author[0000-0002-3900-8208]{Doug Geisler}
\affil{Departamento de Astronomia, Casilla 160-C, Universidad de Concepcion, Chile;
Instituto de Investigaci\'on Multidisciplinario en Ciencia y Tecnolog\'ia, Universidad 
de La Serena. Avenida Ra\'ul Bitr\'an S/N, La Serena, Chile;
Departamento de Astronom\'ia, Facultad de Ciencias, Universidad de La Serena. Av. Juan Cisternas 1200, La Serena, Chile} 

\begin{abstract} 

The distinct stellar populations of the Galactic bulge can be
disentangled through detailed analysis of  their chemical abundances and 
kinematical properties. Recent studies have suggested that
globular clusters located in the Galactic bulge with metallicities around $\rm [Fe/H] \approx -0.7$, may represent some of the oldest systems in the Milky Way, potentially tracing the early spheroidal bulge. The coincidence of a metallicity peak at $\rm [Fe/H] \approx -0.7$ in both field stars and globular clusters, together with the presence of phosphorus-rich (P-rich) stars, may provide important clues to the nature of the
first generations of stars formed in the Galaxy.

In this work, we investigate the odd-Z elements Na, Al and particularly P in the bulge globular clusters NGC~6539 ($\rm[Fe/H] \sim -0.75$) and NGC~6569 ($\rm [Fe/H] \sim -0.85$) using 
APOGEE spectra.
We also examine the clusters Tonantzintla-1 and NGC~6316, which exhibit evidence of phosphorus enhancement.  
Our analysis confirms that NGC~6539 is a cluster of interest, with one clearly P-rich star, whereas NGC~6569 shows a lower level of P-enhancement. This again suggests that there might
have been an early bulge building block with the metallicity of $\rm[Fe/H] \sim-0.75$, of which
NGC~6539 would be part of. The observed abundance patterns indicate that the production of Na and Al is consistent with nucleosynthesis in massive stars. However, the origin of the phosphorus enrichment remains uncertain, suggesting that additional nucleosynthetic channels may be required to explain the observed abundances. These findings provide new constraints on the chemical evolution of the Galactic bulge and the nature of its earliest stellar populations. 
\end{abstract}

\section{Introduction}\label{intro}

The Galactic bulge hosts stellar populations that retain signatures of the Galaxy's earliest assembly history, encoded in their kinematics, orbital properties, chemical abundances, and ages \citep[][and references therein]{chiappini11,barbuy18a,zoccali26}.
However, the present-day bulge is a highly complex structure, making it challenging to disentangle its oldest components 
\citep[see][for recent attempts]{queiroz21, rix22, nepal26}, which may represent some of the earliest stellar populations formed
in the main progenitor of the Milky Way \citep[e.g.][]{geisler25}.
This complexity reflects the coexistence of stellar populations formed both in situ and through accretion events. For example,
\citet{massari26} identified a population of globular clusters in the inner Galaxy that can be linked to the 
accreted structures Heracles \citep{horta21} and Kraken \citep{kruijssen20}.
In contrast, the proto-bulge likely hosted some of the first globular clusters formed in situ within the primordial building blocks of the Milky Way.
The ancient clusters provide a unique fossil record of the earliest phases of bulge assembly and chemical enrichment, and are therefore crucial for understanding the formation of the Galaxy.

Moderately metal-poor globular clusters (GCs) are 
powerful tracers of stellar populations and galaxy formation and evolution, as their properties provide valuable constraints on ages, chemical enrichment histories, and assembly processes
\citep[e.g.][]{souza24a, ortolani26}.
Recent advances in chemical tagging combined with orbital information, have also made possible to associate field stars to their parent globular clusters
\citep[e.g.][]{fernandez-trincado22a,souza24b}.
Particularly relevant in this context are the rare N-rich field stars, which are widely interpreted as former globular cluster members subsequently lost to the field population
\citep[e.g.][]{schiavon17,fernandez-trincado17}, but see \citep[][]{leitinger2026}.

A growing body of evidence suggests the existence of an ancient stellar population in the Galactic bulge characterized by a metallicity peak at $\rm [Fe/H]\approx -0.7$. Three recent results are particularly relevant in this context.

The first important result came from the analysis of large samples of bulge field stars from The Apache Point Observatory Galactic Evolution Experiment \citep[APOGEE;][]{majewski17} combined with radial velocities from Gaia DR3 Radial Velocity Spectrometer \citep[RVS;][]{gaia23}. Using these data,
\citet{nepal26} identified a spheroidal bulge component characterized by pressure-supported kinematics, enhanced $\alpha$-element abundances, and a metallicity distribution peaking at $\rm [Fe/H]\approx-0.7$ dex.

A second key discovery was the identification of P-rich red giant stars. First reported by \citet{masseron20} and subsequently investigated by \citet{brauner23,brauner24}, these stars span the metallicity range $-1.35 < \mathrm{[Fe/H]} < -0.58$ and were found toward the inner disk, bulge, and halo of the Galaxy. The presence of P-rich stars was later confirmed among field stars belonging to the spheroidal bulge by \citet{barbuy25} and \citet{ernandes26}, who found them predominantly within $-1.1 \lesssim \mathrm{[Fe/H]} \lesssim -0.6$. Evidence for phosphorus enhancement was subsequently extended to bulge globular clusters, with P-rich stars detected in Tonantzintla~1 (hereafter Ton~1) 
and NGC~6316, with literature metallicities of $-0.87 < \rm [Fe/H] < -0.61$, and  $-1.00 < \rm [Fe/H] < -0.73$, respectively \citep{barbuy25b}.
The metallicity range where P-rich stars are found overlaps remarkably well with that of the spheroidal bulge population identified by \citet{nepal26}.

The third important result was the age determination of the bulge globular cluster Tonantzintla 2 (hereafter Ton~2). Using deep photometric data,
\citet{ortolani26} derived an age of 13.58$^{+0.72}_{-1.0}$ Gyr for this cluster. Given its metallicity of
$\rm [Fe/H] = -0.7$ dex \citep{fernandez-trincado22b}, Ton~2 was identified as the oldest known globular cluster in the Galactic bulge. This finding further strengthens the possibility that stellar populations with metallicities around this value are associated with the earliest phases of bulge formation.

Taken all together, these findings suggest a possible connection between the ancient spheroidal bulge, moderately metal-poor globular clusters, and the occurrence of phosphorus-enhanced stars. Testing this hypothesis is the main goal of the present work.

In our previous studies we were successful in finding P-enhancement in two GCs: Ton~1 and NGC~6316 both contained several P-rich stars, whereas Ton~2 had a single star with a measured [P/Fe]=+0.8 dex, and three additional stars presented indications of similarly high phosphorus abundances. 
However, in these latter cases the abundance determination relied only on the weaker and less blended line because the stronger line was affected by noise, preventing a firm confirmation of the enhancement
\citep{barbuy25b}. 
As a follow-up of that work, here we analyse APOGEE spectra of the moderately metal-poor GCs NGC~6539 and NGC~6569. These clusters were selected because their metallicities are close to $\rm [Fe/H]\approx-0.7$ dex combined with the availability of APOGEE spectra for member stars. 
By investigating their phosphorus abundances, together with those of other odd-Z elements such as Na and Al, we aim to assess whether phosphorus enhancement is a common characteristic of this ancient bulge population.

NGC~6539 is a relatively poorly studied cluster. High resolution spectroscopic analysis by
\citet{origlia05}, yielded a metallicity $\rm [Fe/H] = -0.76$ and an $\alpha$-enhancement  close to ${\rm [\alpha/Fe]} \sim +0.44$. Other studies based on medium resolution spectroscopy have reported somewhat higher metallicities; for example \citet{dias16} derived $\rm [Fe/H] = -0.55\pm0.06$ from FOcal Reducer and low dispersion Spectrograph 2 at the Very Large Telescope (FORS2/VLT) spectroscopy of eight cluster members.

NGC~6569 on the other hand, is a well-studied cluster. It is 
an outer bulge cluster, as classified by \citet{bica16,bica24}.
The reason for the interest in this cluster is due to the revelation of a possible
double horizontal branch (HB)  detected by \citet{mauro12,mauro14}, based on near-IR photometry from the 
Vista Variables in the Via Lactea \citep[VVV;][]{minniti10}]
survey. The possible dual HB was essentially ruled out by the much higher quality Gemini Multi-conjugate Adaptive Optics System (GeMS) Colour-Magnitude Diagram (CMD) of \citet{saracino19}.
This is a massive cluster with $\rm M = 1.72^{+0.20}_{-0.18}\times10^5 \ M_{\odot}$
\citep{pallanca23}. 

Several spectroscopic studies were carried out for this cluster, many of them with the aim of clarifying if it contains two stellar populations in terms of metallicity, that are found to be in the range  $\rm -0.91  < [Fe/H] < -0.66$ (see Sect. \ref{n6569}).
In terms of CMDs, an early study was carried out by \citet{ortolani01}, and recently by \citet{saracino19} using Hubble Space Telescope (HST)/Wide Field Camera 3 (WFC3) and multi-conjugate adaptive optics assisted GEMINI GeMS/Gemini South data, deriving an age of 12.8$\pm$1 Gyr. Very similar distance moduli of (m-M) = 14.96$\pm$0.2, corresponding to 9.8 kpc, and 15.03$\pm$0.08, corresponding to 10.14 kpc, were given by \citet{ortolani01} and \citet{saracino19}, respectively. 

Finally, the latest estimations of reddening and distance, are E(B-V) = 0.97 for NGC~6539 \citep{piotto02}, and E(B-V)=0.52 \citep{ortolani01,saracino19} for NGC~6569, from 
CMDs. Distances of $8.165^{+0.395}_{-0.379}$, and $10.534^{+0.261}_{-0.255}$ are given by \citet{baumgardt21} for NGC~6539 and NGC~6569, the latter being somewhat higher than the photometric distances.

The manuscript is organized as follows. In Section \ref{sample} the stellar samples are reported.
In Section \ref{n6569} we describe the several spectroscopic analyses of
NGC~6569. In Section \ref{abundance} abundance
derivation of phosphorus and other elements is described. In Section \ref{discussion} the
results are discussed. In Section \ref{conclusions} conclusions are drawn.

\section{The sample}\label{sample}

In this Section we explain why we selected the two sample clusters. The sequence of
facts are: a) the measurement of P in old field bulge stars by \citet{barbuy25}, where we
found high P-abundances in stars with metallicities $\rm [Fe/H]\approx-0.8$, confirming
previous findings by \citet{brauner23}. b)  
\citet{barbuy25b} then inspected the APOGEE-2 clusters NGC~6522 \citep{fernandez-trincado19}, UKS~1 \citep{fernandez-trincado20}, 
Ton~1 or NGC~6380 \citep{fernandez-trincado21c},  and the bulge Cluster APOgee Survey (CAPOS) clusters Ton~2 or Pismis~26 \citep{fernandez-trincado22b}, NGC~6558 \citep{gonzalez23}, HP~1 \citep{henao25} and NGC~6316 \citep{frelijj25}, with some stars in NGC~6380 also observed within CAPOS. 
We found P-rich stars in the clusters Ton-1, and NGC~6316, of metallicities 
around $\rm [Fe/H] \approx -0.7$.
c) In the present paper we verified which GCs observed within the APOGEE survey, given in the compilation by
\citet{schiavon24} would have metallicities close to this.

The selected clusters are projected towards the Galactic bulge, with equatorial coordinates
J2000 $\alpha = 18^{\rm h}04^{\rm m}49.74^{\rm s}$, $\delta = $-$07^{\rm o}35'09.1''$ 
and Galactic coordinates l, b = 20.795$^{\circ}$, 06.776$^{\circ}$ for NGC~6539, and
$\alpha$ = 18$^{\rm h}$13$^{\rm m}$38.88$^{\rm s}$, 
$\delta$ = $-$31$^{\rm o}$49'35.2'', and 
l, b =  0.481$^{\circ}$, -6.681$^{\circ}$ for NGC~6569. 
\citet[][2010 edition]{harris96} reports radial velocities of v$_{\rm r}$ = 31.0 and 
$-$28.1 km.s$^{-1}$ for NGC~6539 and NGC~6569 respectively.

We selected member stars of NGC 6539 and NGC 6569 by applying a proper motion filter, on  the Gaia DR3 \citep{gaia23} 
archive\footnote{https://gea.esac.esa.int/archive/}, to all stars located within the Jacobi radius ($r_{\mathrm{J}}$) of each cluster, 
0.317$^{\circ}$ and 0.226$^{\circ}$ respectively, around their centres \citep{schiavon24}. 
We then cross-matched the resulting sample with APOGEE observations, identifying one star in NGC 6539 and 16 candidate stars in NGC 6569. The final sample is listed in Table \ref{tab:gcpar}, together with stellar parameters, APOGEE radial velocities, and Gaia DR3 proper motions.

\begin{deluxetable*}{@{}ccccccccc@{}}[htbp]

\caption{Sample stars: stellar parameters, radial velocity, signal-to-noise ratio (SNR) from APOGEE, and proper motions from Gaia.}
\label{tab:gcpar}

\tabletypesize{\scriptsize}
\tablehead{\\
APOGEE ID  & $\mathrm{T_{eff}}$ & log g   & $\mathrm{[Fe/H]}$ & $\mathrm{v_{micro}}$ & $\mathrm{v_{r}}$ & $\mu_\alpha$ & $\mathrm{\mu_{\delta}}$ & SNR \\ 
 & \hbox{K} & & & \hbox{km.s$^{-1}$} & \hbox{km.s$^{-1}$}& \hbox{mas.s$^{-1}$} & \hbox{mas.s$^{-1}$} &
}
    \startdata
\multicolumn{8}{c}{ NGC~6569: Members}  \\ 
\midrule
2M18132448-3149140  & 3467$\pm04$ & 0.53$\pm0.03$ & -1.07$\pm0.01 $ & 2.60 & $-49.94 \pm 0.03$ & $-4.16  \pm 0.02$ & $-7.47 \pm 0.02$ & 70 \\
2M18133083-3148103  & 3989$\pm08$ & 1.09$\pm0.04$ & -1.00$\pm0.01$  & 2.30 & $-49.30 \pm 0.03$ & $-3.87 \pm 0.03$  & $-7.49 \pm 0.02$ & 130 \\
2M18133324-3150194  & 4086$\pm08$ & 1.41$\pm0.03$ & -0.88$\pm0.01$  & 1.97 & $-54.96 \pm 0.03$ & $-4.0.6 \pm 0.033$ & $-7.37 \pm 0.02$ & 152 \\
2M18133329-3146211  & 4891$\pm26$ & 2.05$\pm0.05$ & -0.97$\pm0.02$  & 2.49 & $-51.42 \pm 0.10$ & $-4.42 \pm 0.05$ & $-7.71 \pm 0.04$ & 50 \\
2M18133620-3149040  & 4129$\pm09$ & 1.31$\pm0.04$ & -0.96$\pm0.01$  & 2.10 & $-42.58 \pm 0.03$ & $-4.11 \pm 0.03$ & $-7.44 \pm 0.03$ & 16 \\
2M18133789-3147295  & 4084$\pm08$ & 1.22$\pm0.03$ & -1.00$\pm0.01$  & 2.28 & $-50.33 \pm 0.03$ & $-4.20 \pm 0.03$ & $-7.42 \pm 0.02$ & 170 \\
2M18133940-3150132  & 3850$\pm05$ & 0.51$\pm0.03$ & -0.98$\pm0.01$  & 2.39 & $-48.64 \pm 0.03$ & $-4.17 \pm 0.02$ & $-7.48 \pm 0.02$ & 221 \\
2M18134025-3149477  & 3830$\pm06$ & 0.78$\pm0.03$ & -0.99$\pm0.01$  & 2.17 & $-50.45 \pm 0.02$ & $-4.05 \pm 0.10$ & $-7.83 \pm 0.07$ & 174 \\
2M18134151-3148556  & 3955$\pm08$ & 1.17$\pm0.04$ & -0.96$\pm0.01$  & 2.01 & $-55.72 \pm 0.03$ & $-4.1671 \pm 0.03$ & $-7.27 \pm 0.02$ & 119 \\
2M18134725-3147570  & 4539$\pm16$ & 2.22$\pm0.04$ & -0.78$\pm0.01$  & 1.90 & $-46.46 \pm 0.07$ & $-4.23 \pm 0.05$ & $-7.36 \pm 0.04$ & 63 \\
2M18135154-3151406  & 4259$\pm14$ & 1.72$\pm0.04$ & -0.92$\pm0.02$  & 1.83 & $-47.48 \pm 0.06$ & $-3.99 \pm 0.06$ & $-7.50 \pm 0.04$ & 59 \\
2M18142065-3147172  & 4277$\pm13$ & 1.40$\pm0.05$ & -1.04$\pm0.02$  & 2.23 & $-47.16 \pm 0.01$ & $-4.22 \pm 0.04$ & $-7.19 \pm 0.03$ & 91 \\ 
\midrule
\multicolumn{8}{c}{ NGC~6569: Probable non-members}  \\ 
\midrule
2M18140144-3153250  & 3927$\pm05$ & 1.92$\pm0.02$ &  0.43$\pm0.01$  & 1.31 & $-36.08 \pm 0.03$ & $-3.80 \pm 0.04$ & $-6.50 \pm 0.03$ & 94\\
2M18140469-3158520  & 4107$\pm08$ & 1.78$\pm0.03$ & -0.36$\pm0.01$  & 1.59 & $-42.92 \pm 0.03$ & $-4.07 \pm 0.06$ & $-6.37 \pm 0.04$ & 122 \\
2M18132128-3152422  & 4371$\pm12$ & 2.23$\pm0.04$ & -0.43$\pm0.01$  & 1.32 & $-9.75 \pm 0.05$ & $-4.08 \pm 0.04$ & $-7.21 \pm 0.03$ & 70  \\
2M18135591-3153092  & 4425$\pm13$ & 2.28$\pm0.04$ & -0.62$\pm0.01$  & 1.43 & $-1.41 \pm 0.06$ & $-3.95 \pm 0.04$ & $-7.18 \pm 0.03$ & 71  \\
\midrule
\multicolumn{8}{c}{ NGC~6539}  \\
\midrule
2M18042652-0739044  & 4356$\pm15$  & 2.00$\pm0.05$ & -0.74$\pm0.02$   & 1.32 & $30.92 \pm 0.08$ & $-6.88 \pm 0.07$ & $-3.48 \pm 0.06$ & 85
    \enddata
\end{deluxetable*}

\begin{figure}[htp]
    \centering
    \includegraphics[width=0.5\linewidth]{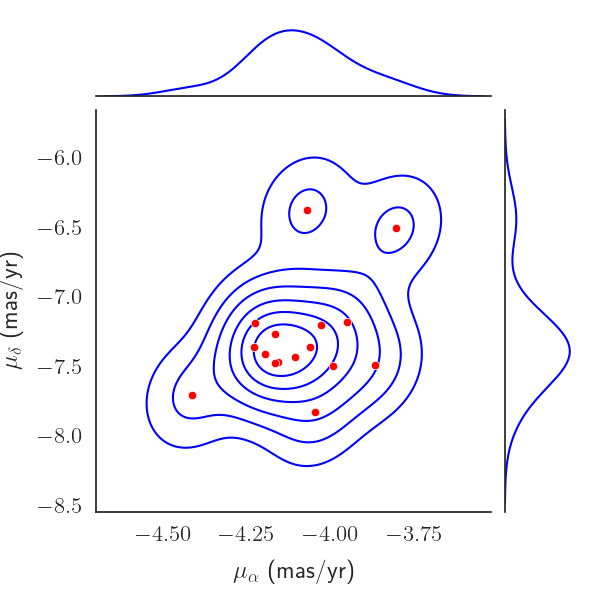}
    \caption{NGC~6569 sample stars proper motion data from Gaia DR3, where it is possible to identify two stars, 2M18140144-3153250 and 2M18140469-3158520, away from the main group located at $\mu_\alpha \approx -4.15$ mas.s$^{-1}$, $\mu_\delta \approx -7.4$ mas.s$^{-1}$.}
    \label{fig:pmrv}
\end{figure}

The proper motion selection allows to filter stars which do not belong to the cluster, identifying visually which data are not clustered, as shown in Figure \ref{fig:pmrv}.
From this test, two stars (2M18140144-3153250 and 2M18140469-3158520) were in principle eliminated as non-members,
for being away from the locus where cluster members are located.

\begin{figure}[htp]
    \centering
    \begin{tabular}{cc}
    \includegraphics[width=0.44\linewidth]{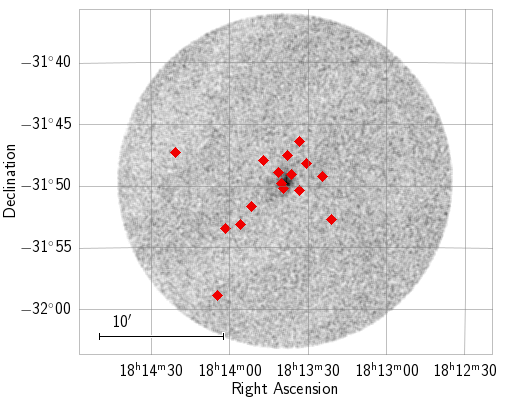}
    \includegraphics[width=0.44\linewidth]{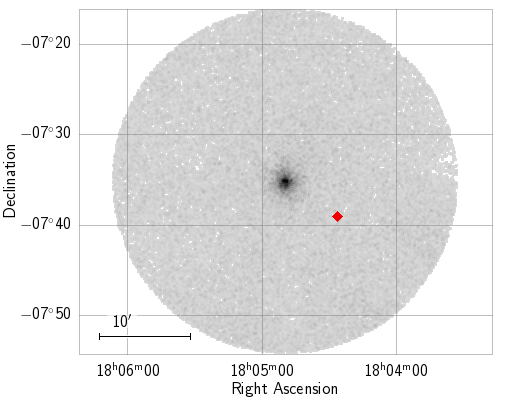}
    \end{tabular}
    \caption{\textit{Left:} NGC~6569 sample stars sky distribution -- red dots are cluster members, grey dots are stars queried from Gaia DR3 with the cluster Jacobi radius of 0.226$^{\circ}$ around its centre. \textit{Right:} Star 2M18042652-0739044 location away from the centre of parent cluster NGC 6539, as a red diamond. Grey dots are stars queried from Gaia DR3 around a radius with the cluster Jacobi radius of 0.317$^{\circ}$ around its centre.}
    \label{fig:6569sky10min}
\end{figure}

From a radial velocity check, the one star in NGC~6539 fits as a member, and among the NGC 6569 stars, 2M18132128-3152422 and 2M18135591-3153092 were excluded, as well from their metallicities. The sky position of NGC 6569 members and NGC 6539 star relative to the cluster are both shown in Figure \ref{fig:6569sky10min}.

\begin{figure}[htp]
    \centering
    \includegraphics[width=1.0\linewidth]{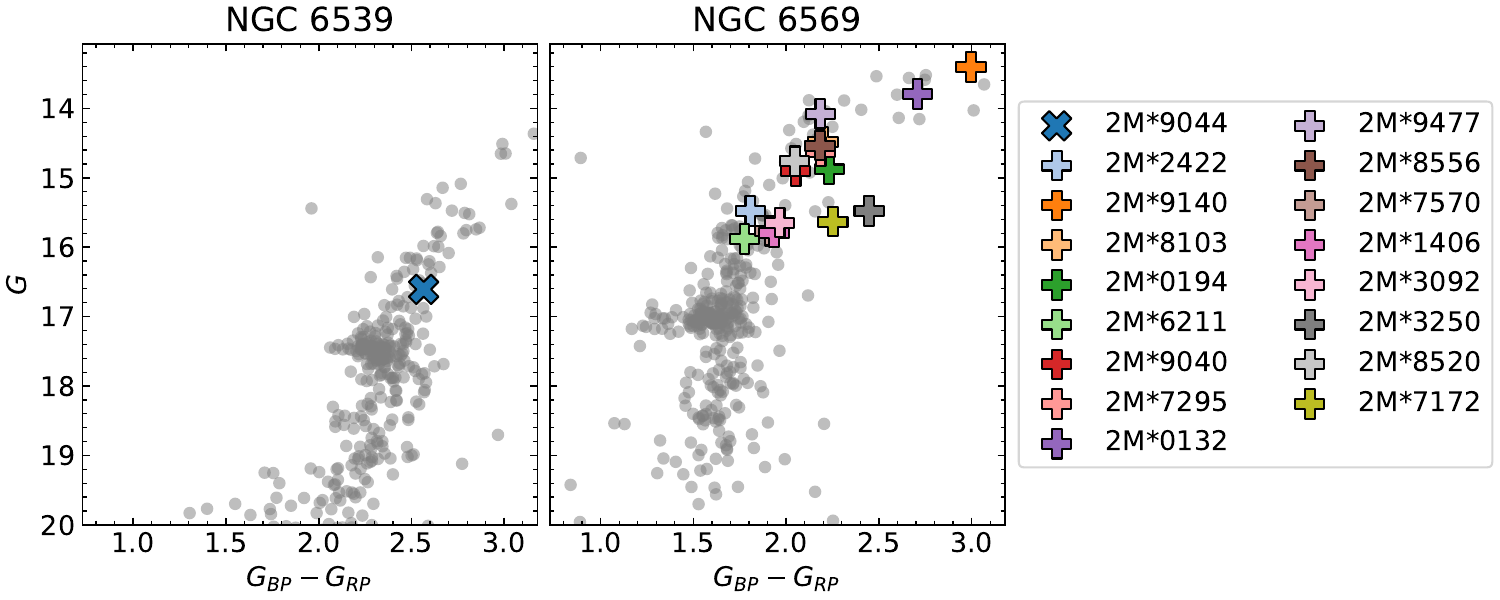}
    \caption{Colour-Magnitude Diagrams of NGC~6539 and NGC~6569 from Gaia data, with the sample stars identified.}
    \label{fig:cmds}
\end{figure}

In Figure \ref{fig:cmds} are shown the CMDs of NGC~6539 and NGC~6569 from Gaia data, with the sample stars identified.

The spectra analysed in the H-band are from the APOGEE survey, part of the Sloan Digital Sky Survey IV \citep[SDSS-IV/V]{blanton17}, having delivered high resolution ($R \sim$ 22,500) and
high signal-to-noise ratios in the \textit{H}-band  (15140-16940 {\rm \AA}) \citep{wilson19}, including  about 7$\times$10$^{5}$ stars.
 APOGEE-1  and APOGEE-2 used the 2.5m Sloan Foundation Telescope at the Apache Point Observatory in New Mexico 
 \citep{gunn06}, and the 2.5m Ir\'en\'ee du Pont Telescope at the Las Campanas Observatory in Chile \citep{bowen73}, respectively. 
 \citet{santana21} and \citet{beaton21} report the target selection
for the Southern and Northern Hemispheres, respectively. 
The detectors are H2RG (2048 x 2048) Near-Infrared HgCdTe Detectors with 18$\mu$ pixels.

The stellar parameter derivation, together with chemical abundances, is carried out through a Nelder-Mead algorithm \citep{nelder65}, with a
simultaneously fit of the stellar parameters effective temperature (T$_{\rm eff}$), gravity (log~g), 
metallicity ([Fe/H]), and microturbulence velocity (v$_{\rm t}$) together with the abundances of carbon, nitrogen, and $\alpha$-elements. The APOGEE Stellar Parameter and Chemical
ASPCAP pipeline \citep{garcia-perez16}, is based on the FERRE code \citep{allende-prieto06} and the APOGEE line list \citep{smith21}.

\section{NGC~6569: a multiple metallicity cluster?}\label{n6569}

The interest in NGC~6569 was triggered by the possible presence of
multiple stellar populations and/or multiple metallicities, deduced
from CMDs by \citet{mauro12,mauro14}. The several studies, 
involving hundreds of stars, are described below.

\begin{itemize}

\item \citet{geisler25} obtained mean [Fe/H] = $-$1.04 $\pm$ 0.05,  [$\alpha$/Fe] = 0.29 $\pm$ 0.08
and v$_{\rm r}$ = $-$49.7 $\pm$3.5 km/s from 9 members with APOGEE spectra from the CAPOS survey.

\item \citet{schiavon24} obtained mean [Fe/H] = $-$0.92 from 14 members with APOGEE spectra.

    \item \citet{valenti11} analysed 6 stars in the H-band, and obtained
[Fe/H] = $-0.79\pm0.02$, and  [$\alpha$-element/Fe] = +0.44. 
    \item \citet{dias16} reports  [Fe/H] = $-0.66\pm0.07$ from FORS2/VLT spectroscopy of 11 stars.
    \item From 148 stars observed in the optical,
\citet{johnson18} obtained a mean metallicity of [Fe/H]=$-0.87\pm0.04$, identifying 
confirmed members by those with radial velocity v$_{\rm r}$ = $-48.8\pm5.3 \ \mathrm{km.s}^{-1}$.
They conclude that there is no sign of a second stellar population of  a different metallicity. 
    \item From 11 stars observed in the H-band, \citet{barrera25} obtained [Fe/H] = $-0.91\pm0.06$.
    \item \citet{hughes26} analysed 303 extra-tidal stars and derived a mean metallicity of
[Fe/H]=$-0.83\pm0.14$, and [$\alpha$-element/Fe] = +0.38. 
    \item From the Multi Unit Spectroscopic Explorer (MUSE/VLT) data, \citet{pallanca23} selected and analysed 485 from 1300 initial targets, resulting in a  v$_{\rm r}$ = $-48.5\pm0.3 \ \mathrm{km.s}^{-1}$ for member stars, covering the Colour-Magnitude Diagram (CMD) from the turn-off to the red giant branch (RGB) tip.
    \item \citet{cohen17} estimated [Fe/H]=$-0.72\pm0.14$ from VVV photometry, and its metallicity
was further discussed in \citet{cohen18} based on HST data.
\end{itemize}

\begin{figure}[htbp]
    \centering
    \includegraphics[width=0.5\linewidth]{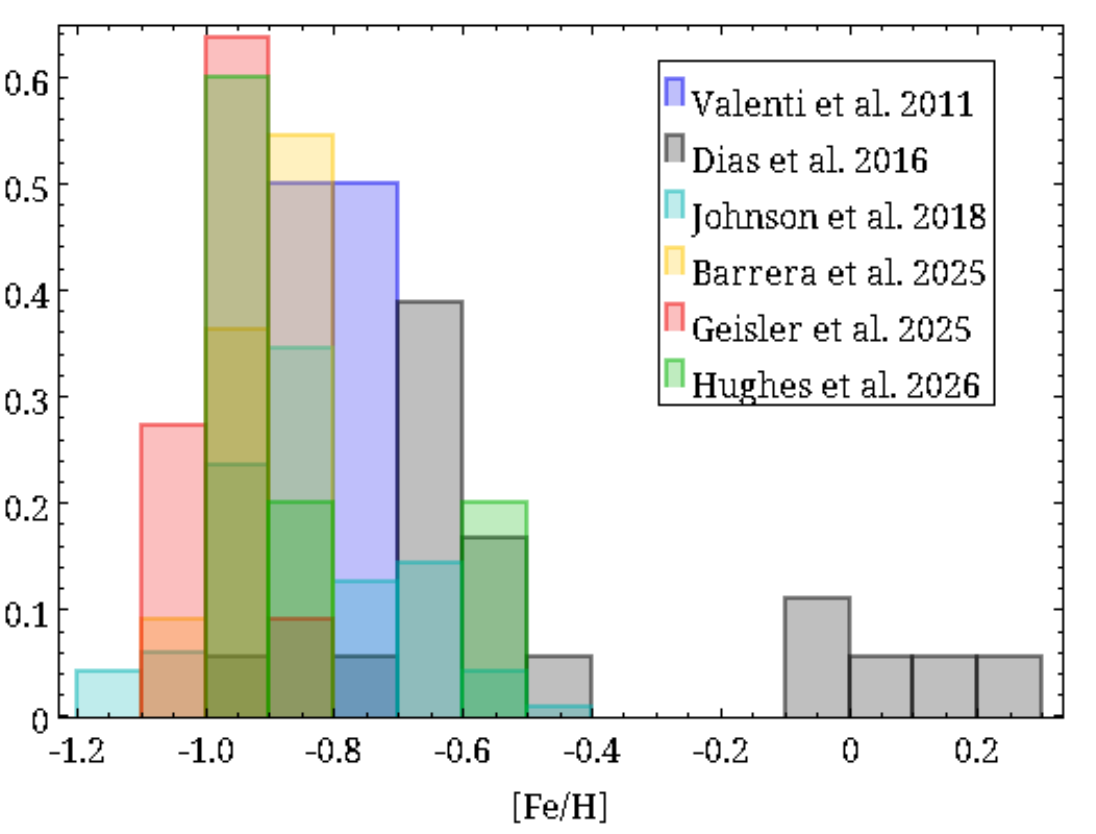}
    \caption{Histogram of metallicity distribution from the literature,
    including results from \citet{valenti11}, \citet{dias16}, \citet{johnson18},
    \citet{barrera25}, \citet{geisler25}, and \citet{hughes26}.}
    \label{fehlit}
\end{figure}

Figure \ref{fehlit} shows the metallicity histograms with samples from \citet{valenti11}, \citet{dias16}, \citet{johnson18}, \citet{barrera25}, \citet{geisler25}, and \citet{hughes26}. It is interesting
to note the large range of metallicities, however 
 there is no  evidence, from each of these studies, for an intrinsic metallicity spread in NGC 6569. We nevertheless analyse four additional stars that are likely non-members, in order to investigate the nature of the stellar populations observed along the line of sight. These objects may help to explain the presence of secondary sequences or contaminating populations superimposed on the cluster CMDs.

\section{Phosphorus abundances}\label{abundance}

In this paper our aim is the derivation of Phosphorus abundances.
For our computations of Phosphorus abundances the spectrum synthesis code \textsc{Turbospectrum} from \citet{alvarez98} and \citet{plez12}, together with Model Atmospheres with a Radiative and Convective Scheme (MARCS) atmospheric models from \citet{gustafsson08} were used. The solar abundance adopted for Phosphorus, as well as Carbon, Nitrogen, Oxygen, Sodium, and Aluminium are from \citet{asplund21}:
A(P) $=$ 5.41, A(C) $=$ 8.46, A(N) $=$ 7.83, A(O) $=$ 8.69,
A(Na) $=$ 6.22, and A(Al) $=$ 6.43.
In Table \ref{linelist} are reported the lines of Phosphorus, Sodium, and Aluminium used
in this work, together with respective oscillator strengths.
The complete atomic line list used is that from the APOGEE collaboration, together with the molecular lines described in \citet{smith21}, namely CO \citep{li15}, OH \citep{brooke16}, CN \citep{sneden14}, C$_2$ \citep{yurchenko18}, FeH \citep{Hargreaves10}, besides TiO \citep{jorgensen94}.

\begin{deluxetable*}{crccccc}[htbp]
    \caption{Line list and oscillator strengths.}
    \label{linelist}
    \tablehead{\\ 
    \colhead{Ion} & \colhead{$\lambda$} & \colhead{$\chi_{ex}$}  &\colhead{log~gf}  &\colhead{log~gf}  &\colhead{log~gf} \\
    & \colhead{(\AA)} &\colhead{(eV)} & \colhead{VALD3}   & \colhead{Kurucz} & \colhead{APOGEE}    
    }
    \startdata
\hbox{PI}     & 15711.522 & 7.176 & $-$0.510 & $-$0.720 & $-$0.404\\
& 16482.932 & 7.213 & $-$0.290 & $-$0.400 & $-$0.273 \\ 
\hbox{NaI} 
& 16388.858 & 3.754 & $-$1.030 & $-$1.030 & $-$1.027 (hfs)& \\
\hbox{AlI} & 16718.957 & 4.085 & 0.290 & 0.152 & 0.220 (hfs)&   \\
& 16750.539 & 4.088 & --- & 0.408 & 0.408 (hfs) &   \\
& 16763.359 & 4.087 & $-$0.524 & $-$0.550 & $-$0.480 (hfs) &   \\
    \enddata
    \tablenotetext{}{Oscillator strengths  from VALD3 \citep{ryabchikova15},  \citet{kurucz95}, 
and the APOGEE collaboration (adopted) are reported. Lines of \ion{Na}{1} and \ion{Al}{1} are split in their hyperfine structure (hfs) components.}
    
\end{deluxetable*}

\begin{figure}[htbp]
    \centering
    \includegraphics[width=0.5\linewidth]{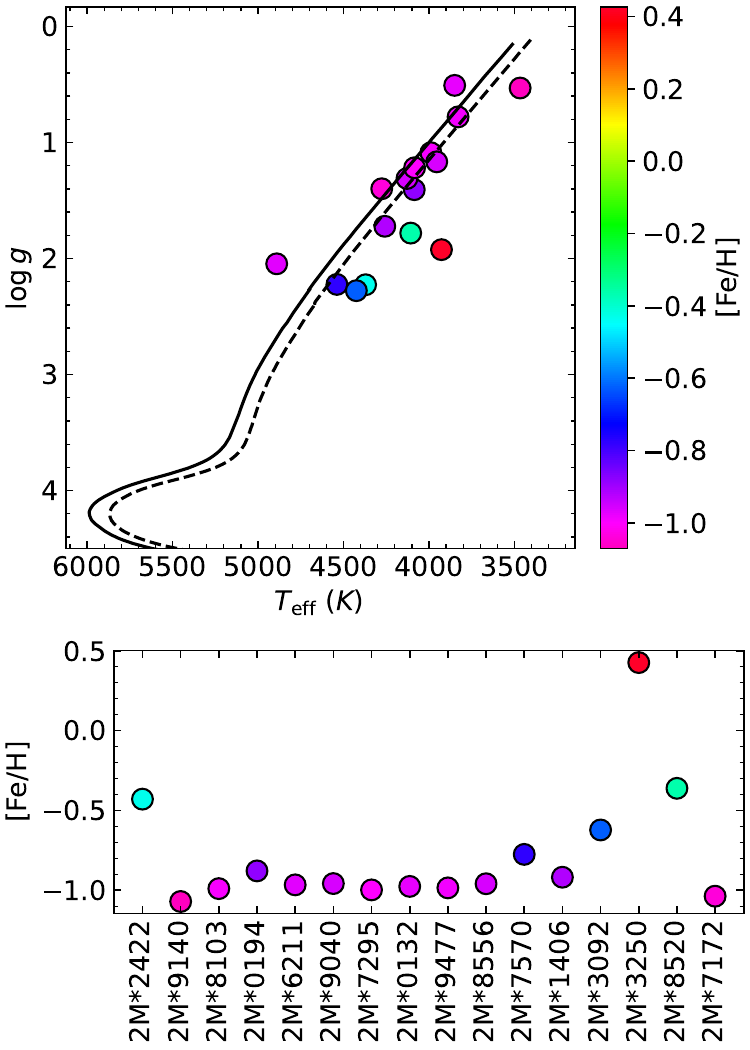}
    \caption{Kiel diagram for NGC~6569 stars, with an isochrone of [Fe/H] = $-$0.87, 
    [$\alpha$/Fe]=0.4 an age of 12.51 Gyr from \citet{saracino19}.}
    \label{fig:kiel}
\end{figure}

The stellar parameters effective temperature (T$_{\rm eff}$), surface gravity (log~g), 
metallicity ([Fe/H]), and microturbulence velocity (v$_{\rm t}$) adopted from the APOGEE ASPCAP 
results \citep{garcia-perez16},
 specifically from their \textsc{Turbospectrum} calculations, are listed in Table \ref{tab:gcpar}. These parameters, resulting from spectroscopic procedure,
are also named uncalibrated, as opposed to the calibrated ones. The use of the uncalibrated parameters is justified because no calibrations were done
for stars cooler than 4100 K, and from our experience in \citet{razera22}, derived abundances
with the calibrated parameters are not suitable. Figure \ref{fig:kiel} plots the stellar parameters in a Kiel Diagram with a Dartmouth Stellar Evolution Database -- DSED alpha-enhanced isochrone, with [alpha/Fe]=+0.4 \citep{dotter08},  and an age of 12.5 Gyr \citep{saracino19}.

The resulting abundances of Carbon, Nitrogen, Oxygen, Phosphorus, Sodium and Aluminium are reported in Table \ref{tab:gcab} for all 16 selected stars of NGC 6569 and the single star of NGC 6539. The abundances of [C/Fe], [N/Fe], and [O/Fe] are crucial, given the blending of a CO feature with the
main stronger \ion{P}{1} 16482.9 {\rm \AA} line.
As described in \citet{barbuy21,razera22,barbuy25}, we first use the region 15525 -- 15590 {\rm \AA}, which contains lines of $^{12}$C$^{16}$O, $^{16}$OH, and $^{12}$C$^{14}$N,  and the strength of CO lines are checked with the line CO 15717.2 {\rm \AA} near the weaker \ion{P}{1} line.

\begin{figure}[htbp]
    \centering
\includegraphics[width=\columnwidth]{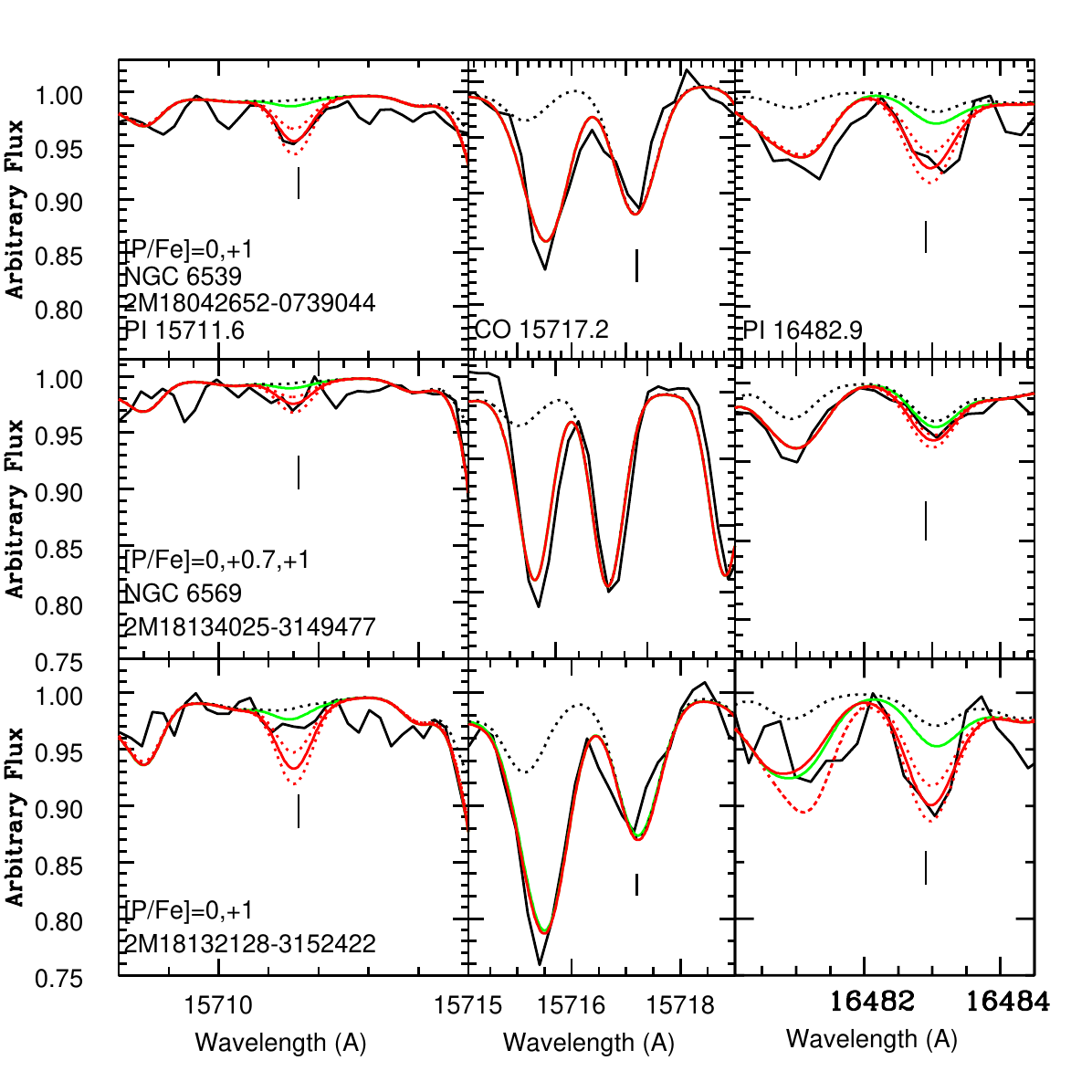}
    \caption{Phosphorus lines fitted with synthetic spectra computed with 
    [P/Fe] = 0.0 
    (solid green lines), and final values (solid red lines), and $\pm$0.2 (dotted red lines) 
    for stars NGC~6539: 2M18042652-0739044, NGC~6569: 
    2M18134025-3149477, and the non-member star
    2M18132128-3152422.}
    \label{fig:phosphorus}
\end{figure}

In Figure \ref{fig:phosphorus} are plotted the 15711.6 and 16482.9 {\rm \AA}
lines, as well as the CO line at 15717.2 {\rm \AA}, compared with synthetic spectra, for
    stars NGC~6539: 2M18042652-0739044, and NGC~6569:
    2M18134025-3149477, as well as the probable non-member star
    2M18132128-3152422, that shows a high Phosphorus abundance.

We also derived Sodium and Aluminium abundances for the sample stars of GCs Ton~1 and NGC~6316, that showed
P-enhancements as reported in \citet{barbuy25b}. The Na lines in both these clusters are strong
and well-defined. We had concluded that the \ion{Na}{1} 16388 {\rm \AA} line is a difficult and not reliable feature in \citet{barbuy23}, for the reason that the Sodium lines were weak in that
more metal-poor sample, and disturbed by blends with molecular lines,
besides noise.
This is not the case in the present sample,
for which the Sodium lines are strong, and with good SNR.
Also the stars are not in the radial velocity
range of -110 to -6 km.s$^{-1}$, where a sky line coincides with this Sodium line \citep{hayes22}.
The 3 Aluminium lines in the H-band are clearly measurable in all
these samples, where the 2nd and 3rd lines are more sensitive to the Aluminium abundance,
therefore more reliable than the 1st line. This said, all 3 lines are well-fitted in most cases.
These Sodium and Aluminium abundances in Ton~1 and NGC~6316 are reported in Appendix A.

\subsection{Uncertainties}

In Table \ref{tab:gcab} are given the uncertainties due to fitting uncertainties, including noise,
continuum placement, and sensitiveness of the line to the abundance.

\begin{deluxetable*}{@{}ccccccccccl@{}}[htbp]
  \caption{Abundances derived for C, N, O,  Na, Al, and P for sample stars.}
   \label{tab:gcab}
  \tabletypesize{\fontsize{4}{8}\selectfont }
  \setlength{\tabcolsep}{2.5pt}

    \tablehead{
    \\\\
    APOGEE ID & \hbox{[Fe/H]} & \hbox{[C/Fe]} & \hbox{[N/Fe]} & \hbox{[O/Fe]} &  [Na/Fe] & [Al/Fe] & \hbox{[P/Fe]}     & \hbox{[P/Fe]}    & \hbox{[P/Fe]}   & comments on P lines\\ 
&          &               &               &               &          &         & $\lambda$15711.2  & $\lambda$16480.2  & mean           &
    }
    \startdata
\multicolumn{11}{c}{ NGC~6569: Members}  \\ \hline
2M18132448-3149140  & $-$1.07 & +0.20$\pm$0.05 & +0.25$\pm$0.15 & +0.65$\pm$0.15 & +0.40$\pm$0.05 & +0.09$\pm$0.15 &---   & ---  & ---   & noise and APOGEE gap         \\
2M18133083-3148103  & $-$1.00 & +0.15$\pm$0.05 & +0.40$\pm$0.10 & +0.47$\pm$0.10 & +0.35$\pm$0.05 & +0.20$\pm$0.05 & ---  & ---  & ---   & noise and APOGEE gap      \\
2M18133324-3150194  & $-$0.88 & +0.12$\pm$0.05 & +0.30$\pm$0.10 & +0.50$\pm$0.10 & +0.10$\pm$0.10 & +0.30$\pm$0.05 &+1.0$\pm$0.20 &+0.0$\pm$0.20  &+0.5$\pm$0.20   & uncertain     \\
2M18133329-3146211  & $-$0.97 & +0.25$\pm$0.25 & +0.30$\pm$0.25 & +0.50$\pm$0.25 & +1.00$\pm$0.15 & +1.10$\pm$0.10 &$>$1.0$\pm$0.40& ---  &$>$1.0$\pm$0.40 & uncertain     \\
2M18133620-3149040  & $-$0.96 & +0.15$\pm$0.05 & +0.25$\pm$0.10 & +0.40$\pm$0.10 & +0.00$\pm$0.10 & +0.30$\pm$0.05 & ---  &+0.35$\pm$0.20 & +0.35$\pm$0.20 & shallow line, uncertain        \\
2M18133789-3147295  & $-$1.00 & +0.15$\pm$0.05 & +0.20$\pm$0.05 & +0.45$\pm$0.10 & +0.30$\pm$0.10 & +0.25$\pm$0.05 & ---  & ---  & ---   &     \\
2M18133940-3150132  & $-$0.98 & -0.15$\pm$0.05 & +0.70$\pm$0.05 & +0.60$\pm$0.10 & +0.80$\pm$0.05 & +0.90$\pm$0.05 & ---  & +0.5$\pm$0.20 & +0.5$\pm$0.20  & shallow line, uncertain \\
2M18134025-3149477  & $-$0.99 & +0.30$\pm$0.05 & +0.25$\pm$0.05 & +0.72$\pm$0.05 & +0.35$\pm$0.10 & +0.30$\pm$0.05 & +0.7$\pm$0.10 & +0.7$\pm$0.10 & +0.7$\pm$0.10  & both lines agree        \\
2M18134151-3148556  & $-$0.96 & +0.15$\pm$0.05 & +0.25$\pm$0.10 & +0.40$\pm$0.10 & +0.25$\pm$0.10 & +0.25$\pm$0.10 & ---  & ---  & ---   & shallow P line      \\
2M18134725-3147570  & $-$0.78 & +0.20$\pm$0.08 & +0.30$\pm$0.08 & +0.35$\pm$0.08 & +0.60$\pm$0.10 & +0.60$\pm$0.05 &---   & +0.5$\pm$0.20 &  +0.5$\pm$0.20 & uncertain \\
2M18135154-3151406  & $-$0.92 & $-$0.10$\pm$0.05 & $-$0.32$\pm$0.15 & $-$0.20$\pm$0.15 & +0.00$\pm$0.10 &$-$0.20$\pm$0.05 & --- & +0.5$\pm$0.10 & +0.5$\pm$0.10  &      \\
2M18142065-3147172  & $-$1.04 & +0.15$\pm$0.05 & +0.25$\pm$0.05 & +0.40$\pm$0.10 & +0.55$\pm$0.15 & +0.40$\pm$0.15 & ---  & +0.6$\pm$0.20 & +0.6$\pm$0.20  &      \\
\midrule
\multicolumn{11}{c}{ NGC~6569: Probable non-members}  \\ 
\midrule
2M18140144-3153250 & $+$0.43 & +0.50$\pm$0.10 & +0.70$\pm$0.10 & +0.70$\pm$0.10 &+1.00$\pm$0.05 & +0.65$\pm$0.10 & +1.0$\pm$0.20 & +0.7$\pm$0.20 & +0.85$\pm$0.30 & strong CO in line 2      \\
2M18140469-3158520 & $-$0.36 & +0.10$\pm$0.05 & +0.15$\pm$0.10 & +0.20$\pm$0.05 &+0.05$\pm$0.05 & +0.25$\pm$0.05 & ---  & ---  & ---  &     \\
2M18132128-3152422 & $-$0.43 & +0.15$\pm$0.08 & +0.20$\pm$0.08 & +0.30$\pm$0.10 &+0.40$\pm$0.15 & +0.15$\pm$0.05 & +0.8$\pm$0.20 &+1.0$\pm$0.10  & +0.9$\pm$0.15 & noise in line 1 \\
2M18135591-3153092 & $-$0.62 & +0.20$\pm$0.05 & +0.20$\pm$0.15 & +0.40$\pm$0.15 &+0.10$\pm$0.15 & +0.10$\pm$0.10  & ---  &+0.8$\pm$0.20  & +0.8$\pm$0.20 & noise \\
\midrule
\multicolumn{11}{c}{ NGC~6539}  \\
\midrule
2M18042652-0739044 & $-$0.74 & +0.08$\pm$0.05 & +0.15$\pm$0.10 & +0.40$\pm$0.10 &+0.20$\pm$0.10 & +0.30$\pm$0.05 & +1.0$\pm$0.05 & +1.0$\pm$0.05 & +1.0$\pm$0.05 & $a$ \\ 
    \enddata
    \tablenotetext{a}{The abundances of Phosphorus are derived from two different lines, and the mean abundance is also given. The reported uncertainties refer to quality of fit, including continuum level, and noise.}
\end{deluxetable*}

Uncertainties from stellar parameters on the Phosphorus abundances, that are negligible for a warm star,
are more significant for the cooler stars, as given in the Appendix.

Another approach comes from  a cross-match with the 11 stars analysed by \citet{barrera25}, in common with our sample, as this cluster was included in the The bulge Cluster APOgee Survey \citep[CAPOS;][]{geisler25}.
Differences between ASPCAP and \citet{barrera25}'s adopted photometric 
parameters are: 
$\rm {T_{ eff}}=^{+152}_{-66}$ K. $\rm [Fe/H]=^{+0.06}_{-0.09}$, $\rm \Delta v_{t}=+0.57$ \citep[ASPCAP][]{barrera25}, and $\rm \Delta log~g=+0.25$ \citep[ASPCAP][]{barrera25}. Five stars are in common with \citet{johnson18}, and the stellar parametermean differences are $\rm \Delta T_{eff} = -138$ \citep[ASPCAP][]{johnson18}. $\rm \Delta [Fe/H]= 0.02$ \citep[ASPCAP][]{johnson18}, $\rm \Delta$v$_{t} = +0.16$ \citep[ASPCAP][]{johnson18}, and $\rm log~g=\pm0.08$.

Relative to ASPCAP abundances, there are differences in Carbon, Nitrogen and Oxygen abundances, but lines of CO, OH and CN are well-fitted. Some differences in Sodium are probably due to noise and misleading fits by ASPCAP,
whereas ASPCAP Aluminium abundances are similar to ours. Phosphorus abundances are not available in ASPCAP.

Finally, uncertainties due to deviations from the Local Thermodynamic Equilibrium (non-LTE), are now available, thanks to the recent work by \citet{andrievsky26}. It is shown that non-LTE effects are small. As can be seen from their Table A.3, for $\rm [Fe/H]=-0.5$, and effective temperatures below 4500 K, corrections on the \ion{P}{1} 16482.9 {\rm \AA} line are smaller than $-$0.1 dex.

\section{Discussion}\label{discussion}

 This paper is a follow-up of the analysis presented in \citet{barbuy25b}, where we investigated phosphorus abundances in a sample of bulge globular clusters. P enrichment was previously identified in Ton~1 and NGC 6316, whereas no clear evidence for P-rich stars was found in HP~1, NGC~6522, UKS~1, NGC~6558, and Ton~2, although in the latter two cases the spectra are affected by relatively low signal-to-noise ratios, which prevented a firm exclusion of weak P enhancements.

In the present work, we find that the single analyzed star in NGC 6539, whose metallicity is close to the [Fe/H]=$-$0.7 dex peak associated with the spheroidal bulge, is phosphorus-rich. In NGC 6569, the situation is more mixed:
one star has [P/Fe]=+0.7$\pm$0.10, another two ones have
[P/Fe]=+0.5$\pm$0.10,   and [P/Fe]=+0.6$\pm$0.20. We discarded two
other stars showing [P/Fe]=+1.0$\pm$0.15 but deduced only from the weaker less reliable line.
The predictions of the chemical evolution model reaches [P/Fe] of +0.45 at [Fe/H] around $-$0.85 \citep{barbuy25}, therefore only [P/Fe]$\geq$0.7 can be considered enriched.
Note that for the non-member stars, at least one of them is clearly enhanced,
but we will not discuss further the cases of the non-member stars.

\begin{figure*}
	\centering
	\includegraphics[width=15.5cm]{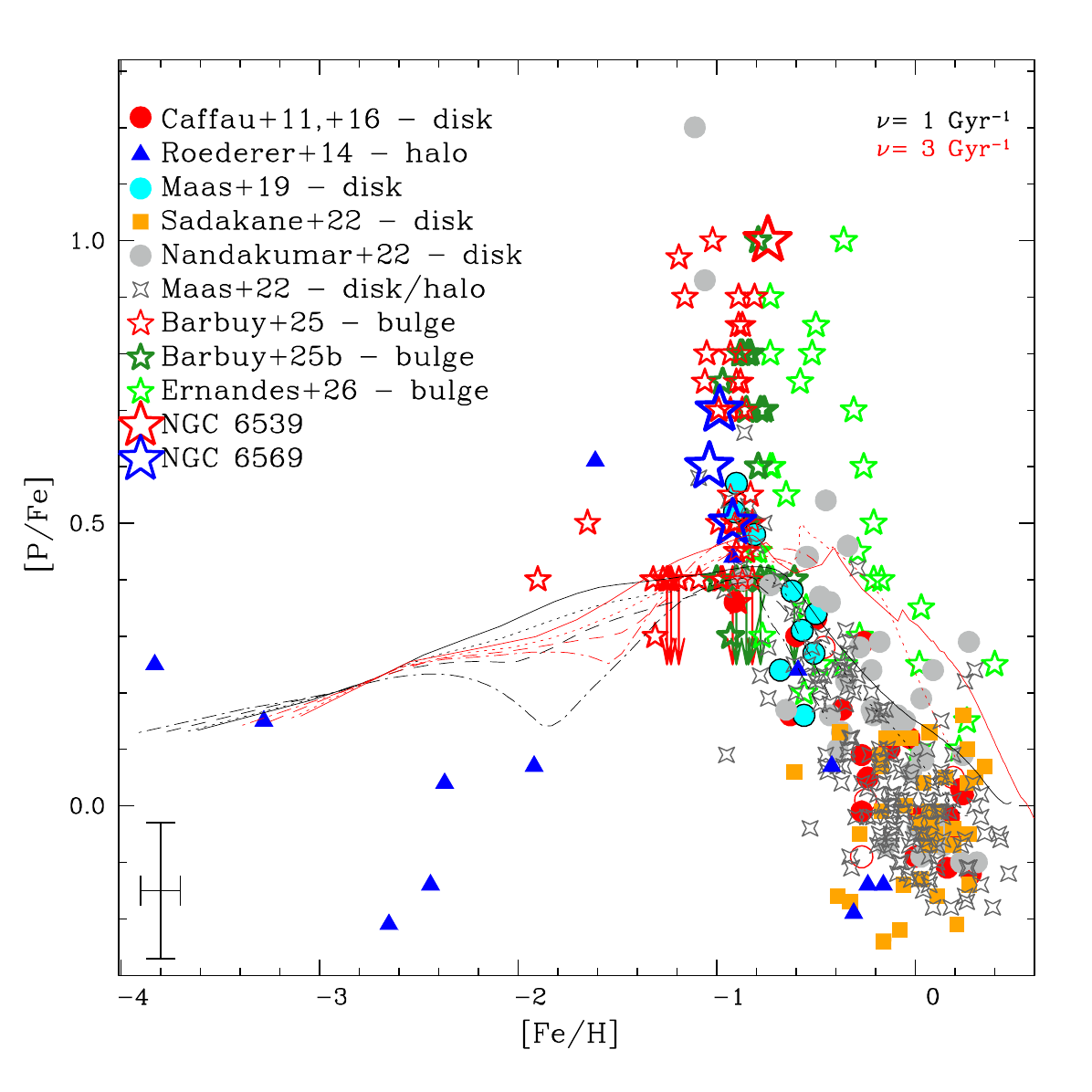}
	\caption{[P/Fe] vs. [Fe/H] for the present results compared with literature data.
		Symbols -- Present work:
        Large red open star: 1 star in NGC~6539; Large blue stars: stars in NGC~6569
        with [P/Fe]=+0.7,+0.6 and +0.5. Literature data:
        red-open stars: \citet{barbuy25},
        darkgreen-open stars: \citet{barbuy25b}, 
        green-open stars: \citet{ernandes26},
		red-filled circles: \citet{caffau11}, red- open circles: \citet{caffau16},
		blue-filled triangles: \citet{roederer14}, 
        filled-cyan circles: \citet{maas19}; 
		light grey-filled circles: \citet{nandakumar22}, 
		light grey open 4-side stars: \citet{maas22}.
		Different model lines correspond to the outputs of models computed for radii r $<$ 0.5, 0.5 $<$ r $<$ 1, 1 $<$ r $<$ 2, 
		and 2 $<$ r $<$ 3 kpc from the Galactic centre. Black lines correspond to
		specific star formation $\nu$ = 1 Gyr$^{-1}$, red lines to $\nu$ = 3 Gyr$^{-1}$.}
	\label{plotpfe}
\end{figure*}

Figure \ref{plotpfe} shows the present results compared with data from the literature and chemical evolution models from \citet{barbuy25}. Previous work include 
\citet{caffau11,caffau16} for disk dwarf stars, \citet{roederer14} in 14 halo stars, \citet{maas19,maas22} in a large sample of disk and halo stars, \citet{sadakane22} in 45 main-sequence stars, \citet{nandakumar22} in 38 disk stars, as well as our previous results from \citet{barbuy25}, \citet{barbuy25b}, and \citet{ernandes26}.

The chemical-evolution model for phosphorus in the Galactic bulge was presented in
\citet{barbuy25}. It consists of a chemical evolution model coupled with hydrodynamic equations,
as first described in \citet{friaca98}. 
We solve the fluid equations of mass, momentum and energy conservation, taking into account the sink terms for the gas due to star formation and source and heating terms due to the late stages of star evolution (supernovae, planetary nebulae, Asymptotic Giant Branch (AGB) stars,  and stellar winds). Also the cooling function of the gas consistently considers the metallicity of the gas.
Models were computed for radii of r $< 0.5$, $0.5 <$ r $< 1$, $1 <$ r $< 2$,
and $2 <$ r $< 3$ kpc from the Galactic centre, and for specific star formation rate
values of $\nu = 1$ and $3$ Gyr$^{-1}$, these values resulting from the good fitting of
our models to the data \citep{friaca17,barbuy25}.

For the nucleosynthesis yields, we adopted
(i) for massive stars,
the metallicity-dependent yields from CCSNe/SNe II from \citet{woosley95}, 
where  the yields from WW95 for sub-solar metallicities were multiplied by a factor of 2 for P, likewise adopted by  \citet{cescutti12}.
At low metallicities (Z $<$ 0.01 Z$_{\odot}$ , or [Fe/H] $< -$2.5), the yields from high-explosion-energy hypernovae from \citet{nomoto13} are employed;
(ii) type Ia supernovae (SNIa) yields from \citet{iwamoto99} – their models W7 
(progenitor star of initial metallicity
Z = Z$_{\odot}$) and W70 (zero initial metallicity); and (iii) for intermediate-mass stars (0.8–8 M$_{\odot}$) with initial Z = 0.001, 0.004, 0.008, 0.02, and
0.4, yields from \citet{vandenhoek97} with variable $\eta$ (AGB case), where
$\eta_{AGB}$ is the mass-loss efficiency of the Reimers' law for AGB mass loss; here we adopt $\eta_{AGB}$ decreasing with metallicity.
In the case of phosphorus, the model considers as well neutrino-interaction processes, which could be important at very low metallicities, to produce odd-Z elements, such as phosphorus, with yields from \citet{yoshida08}, but these effects
are negligible at the moderate metallicity of the studied clusters.


The models reproduce well the general behaviour of P-abundances vs. metallicity.
Overall, the detection of P-rich stars in NGC~6539 and at a lower level
in some stars of NGC~6569 supports the emerging picture that moderately metal-poor bulge globular clusters, particularly those clustering around [Fe/H] = $-$0.7, preserve signatures of early chemical enrichment in the Galactic bulge. This behaviour suggests that the first generations of massive stars in the proto-bulge were efficient producers of odd-Z elements such as Na, Al, and P, although the relative contribution of different nucleosynthetic channels remains uncertain.
The fact that some stars are P-enriched and others not in NGC~6569 may be
related to the same phenomenon of N-rich and N-normal stars in field globular clusters stars \citep[e.g.][]{qiao26},
interpreted as first generation (N-normal) and second generation (N-rich).

\begin{figure}[htbp]
    \centering
\includegraphics[width=\columnwidth]{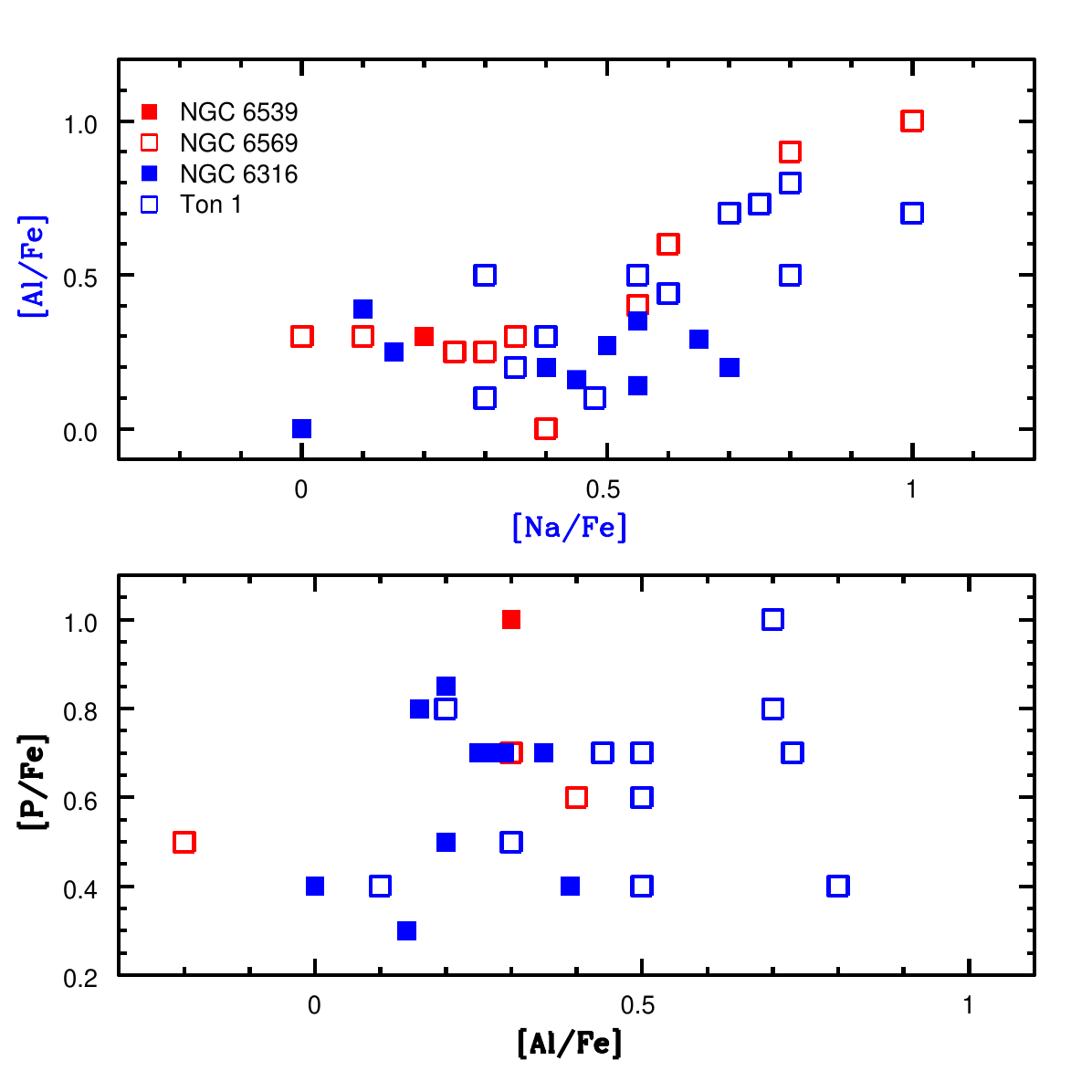}
    \caption{[Al/Fe] vs. [Na/Fe] and [P/Fe] vs. [Al/Fe] for the four clusters:
NGC~6539 red filled square, NGC~6569: red open squares, NGC~6316: blue filled squares, and Ton~1: blue open squares. }
    \label{plotcorrel}
\end{figure}

To explore the chemical connections among these elements, Figure \ref{plotcorrel} presents [Al/Fe] vs. [Na/Fe] and [P/Fe] vs. [Al/Fe] for NGC~6539, NGC~6569, NGC~6316, and Ton~1. A clear Na–Al correlation is observed, consistent with the well-known multiple-population patterns in globular clusters and their dependence on initial cluster metallicity. In contrast, phosphorus does not follow so clearly the same trends, indicating that its production is not simply tied to the processes responsible for Na and Al variations.


Indeed, P is expected to be less sensitive to metallicity than either Na or Al, due to weak interactions that change the neutron excess during carbon and neon burning 
\citep{woosley95}. Enrichment by AGB
stars are in principle excluded \citep{karakaslattanzio14,pignatari16}.
and massive stars should be the object that enriches
the environment in P. The fact that some stars are P-rich and not others
is the same that is found for other elements in
 second generation stars in globular clusters.

The origin of the P enhancement therefore remains uncertain. While massive stars and core-collapse supernovae are the primary candidates,  metallicity-dependent effects in  metal-rich massive-star yields need to be further inspected. In addition, our new data for NGC~6539 and NGC~6569 do not confirm the correlation between [P/Fe] and [N/O] reported in \citet{barbuy25b}.



\section{Conclusions}\label{conclusions}

We have measured abundances of P, as well as of Na and Al, in the two moderately metal-poor globular clusters NGC~6539, and NGC~6569,
that had available spectra observed in the H-band, from the APOGEE survey.

This study was motivated by the fact that \citet{ortolani26} found that the moderately metal-poor globular cluster Tonantzintla~2,
with [Fe/H]$\sim$ $-$0.7 dex,
is the oldest cluster in the Galactic bulge, and older than the more well-known moderately metal-poor ([Fe/H]$\sim$ $-$1.0) clusters
HP~1, NGC~6522, NGC~6558, among others \citep[see][]{souza24a}. Besides, another new revelation was given by \citet{nepal26},
that studied large samples from APOGEE and Gaia
Radial Velocity Spectrometer (RVS) data, and concluded that stars from an original spheroidal bulge 
have a metallicity peak at [Fe/H]$\sim$ $-$0.7.
A third point is the enhancement in phosphorus in several field and globular cluster stars around this same metallicity.

We find that the unique star available in APOGEE data of NGC~6539 is P-rich, 
with [P/Fe]=+1.0$\pm$0.1. The case of NGC~6569 is less clear:
one star has [P/Fe]=+0.7$\pm$0.10, and another two have [P/Fe]=+0.5$\pm$0.10, and 
 [P/Fe]=+0.6$\pm$0.20. We have not considered the results deduced from the weaker P line.
This supports the idea P-rich stars are less frequent at
metallicities lower than [Fe/H]$<$$-$0.7. 

As final conclusions, we suggest that NGC~6539 is a cluster of interest for further studies.
We also suggest that a now disrupted 
massive building block having reached a metallicity of [Fe/H]$<$$-$0.7 was present in the early Galactic bulge, having given birth to P-rich stars.

 We also find a correlation between Na and Al abundances, but not a clear one between P and Al. For the relation between [P/Fe] vs. [N/O] found in \citet{barbuy25b}, we found no further evidence confirming this effect. 

In conclusion, the earliest core-collapse supernovae,  may have been key in producing odd-Z elements (Na, Al, P) in great quantities in the early stages of the Galaxy, as we can observe in the moderately metal-poor globular clusters, important tracers for the early spheroidal bulge formation. The mechanism of P production remains however unclear and calls urgently for the development of stellar models of massive stars in the intermediate/metal-rich range.

This study provides support to the findings of P-rich stars in
an old bulge stellar population with metallicity of $\rm [Fe/H]\approx-0.7$
that we interpret as evidence suggesting a particular
early building block in the Galactic bulge.

\begin{acknowledgments}
We are grateful to Stan Woosley for useful comments on the production of phosphorus
in massive stars.
M.S.C., B.B., A.C.S.F., E.B. and C.C acknowledge grants from FAPESP, Conselho Nacional de Desenvolvimento Cient\'ifico e Tecnol\'ogico (CNPq) and Coordena\c{c}\~ao de Aperfei\c{c}oamento de Pessoal de N\'ivel Superior (CAPES) - Financial code 001. 
H.E. acknowledges a post-doctoral fellowship at Lund Observatory.
SOS acknowledges the DGAPA–PAPIIT grant IA103224 and the support from Dr. Nadine Neumayer's Lise Meitner grant from the Max Planck Society.
J.G.F-T gratefully acknowledges the support provided by ANID Fondecyt Regular No. 1260371, the Joint Committee ESO-Government of Chile under the agreement 2023 ORP 062/2023 and the support of the Doctoral Program in Artificial Intelligence, DISC-UCN.
N.B. acknowledges support from ANID / Scholarship Program / DOCTORADO NACIONAL / 2025 - 72180000.
D.G. acknowledges financial support from the Vicerrector\'ia de Investigaci\'on y Postgrado de la Universidad de La Serena.
\end{acknowledgments}

\bibliographystyle{apj}{}
\bibliography{references}

\begin{appendix}

\section{Na and Al abundances in Ton~1 and NGC~6316}

In Table \ref{ton1} are reported abundances for the
clusters Ton~1 and NGC~6316.
Na abundances are derived in this work. P, N, and O were derived in \citet{barbuy25b}, and Mg, Al were derived in
\citet{fernandez-trincado21c}, and \citet{frelijj25}.

\begin{table}
\centering
\caption{Abundances of P, Mg, Na, Al for Tonantzintla~1 and
NGC~6316. }
\label{ton1}
\small
\scalefont{0.9}
\begin{tabular}{lccccc@{}@{}}
\toprule
\hbox{Star} & \hbox{[P/Fe]} & \hbox{[N/O]} & \hbox{[Mg/Fe]} & \hbox{[Na/Fe]} & \hbox{[Al/Fe]} \\  
\midrule
Ton1:2M173429213904514     &   +0.60 &   +0.60 &  +0.20  & +0.55  & +0.5  \\
Ton1:2M173436163903344     &   +0.50 &   +0.45 &  +0.33  & +0.40  & +0.3  \\
Ton1:2M173430253903190     &   +0.70 &   +0.67 &  +0.31  & +0.60  & +0.44 \\
Ton1:2M173426933904060     &   +0.40 &   +0.25 &  +0.27  & +0.30  & +0.1  \\
Ton1:2M173425413902338     &   +0.40 &   +0.33 &  +0.33  & +0.80  & +0.8  \\
Ton1:2M173427673903405     &   +1.00 &   +0.65 &  +0.36  & +1.00  & +0.7  \\
Ton1:2M173419693905457     &   +0.80 &   +0.75 &  +0.32  & +0.70  & +0.7  \\
Ton1:2M173421773906173     &   +0.70 &   +0.80 &  +0.31  & +0.80  & +0.5  \\
Ton1:2M173429433902500     &   +0.70 &   +0.50 &  +0.32  & +0.75  & +0.73 \\
Ton1:2M173435213903091     &   +0.80 &   +0.80 &  +0.22  & +0.35  & +0.2  \\
Ton1:2M173425883901406     &   +0.40 &   -0.30 &  +0.18  & +0.48  & +0.1  \\
Ton1:2M173419223906052     &   +0.40 &   -0.65 &  +0.48  & +0.30  & +0.5  \\
NGC6316:2M171638642809385  &   +0.70 &   -0.20 &  +0.27  & +0.65  & +0.29 \\
NGC6316:2M171636232808067  &   +0.70 &   +0.65 &  +0.03  & +0.50  & +0.27 \\
NGC6316:2M171633302808396  &   +0.80 &   +0.90 &  +0.34  & +0.45  & +0.16 \\
NGC6316:2M171640482808443  &   +0.70 &   +0.25 &  +0.32  & +0.55  & +0.35 \\
NGC6316:2M171644822808302  &   +0.70 &   -0.30 &  +0.39  & +0.15  & +0.25 \\
NGC6316:2M171633932811052  &   +0.50 &   +0.10 &  +0.35  & +0.70  & +0.2  \\
NGC6316:2M171639032807212  &   +0.85 &   +0.5  &  +0.16  & +0.40  & +0.2  \\
NGC6316:2M171652352809502  &   +0.40 &   -0.55 &  +0.39  & +0.10  & +0.39 \\
NGC6316:2M171639112804506  &   +0.40 &   -0.43 &  +0.30  & +0.00  & +0.0  \\
NGC6316:2M171636272807166  &   +0.30 &   +0.00 &  +0.32  & +0.55  & +0.14 \\
\bottomrule
\end{tabular}
\end{table}

\section{uncertainties}

Typical uncertainties are computed by adopting errors in the stellar parameters of $\Delta$T$_{\rm eff}$ = 100 K, $\Delta$log \textit{g} $=$ 0.2, $\Delta$v$_{\rm t} = -$0.3 km s$^{-1}$, shown in Table \ref{errors}, applied to the
warm star NGC~6539:2M8042652-0739044, and the cool  star NGC~6569-2M18134025-3149477.
The CNO abundances were first adapted to the modified stellar parameters, which is
important due to the blend of the main PI line with CO.

\begin{table}
\caption{Abundance uncertainties due to stellar parameters for the
warm star NGC~6539:2M8042652-0739044,
and the cool  star NGC~6569-2M18134025-3149477,
 for uncertainties of $\Delta$T$_{\rm eff}$ = 100 K,
$\Delta$log g = 0.2, $\Delta$v$_{\rm t}$ = 0.3 km s$^{-1}$ and
corresponding total error.} 
\label{errors}
\centering
\begin{tabular}{lcccccccccccc}
\noalign{\smallskip}
\hline
\noalign{\smallskip}
\hline
\noalign{\smallskip}
\hbox{Element} & \hbox{$\Delta$T} & \hbox{$\Delta$log $g$} & 
\phantom{-}\hbox{$\Delta$v$_{t}$} & \phantom{-}\hbox{($\sum$x$^{2}$)$^{1/2}$} &\\
\hbox{} & \hbox{100 K} & \hbox{0.2 dex} & \hbox{0.3 kms$^{-1}$} & & \\
\hbox{(1)} & \hbox{(2)} & \hbox{(3)} & \hbox{(4)} & \hbox{(5)}  &\\
\hline
\hline
\noalign{\smallskip}
\hbox{} & \hbox{} & \hbox{NGC~6539:2M8042652-0739044} & \hbox{} & \hbox{}  &\\
\noalign{\hrule\vskip 0.1cm}
\hline
\noalign{\smallskip}
\hbox{[C/Fe]}      &   0.06 &  0.05 & 0.00 & 0.08 & \\
\hbox{[N/Fe]}      &   0.10 &  0.10 & 0.00 & 0.14 & \\
\hbox{[O/Fe]}      &   0.15 &  0.05 & 0.02 & 0.16 & \\
\hbox{[Na/Fe]}     &  0.03  &  0.00 & 0.00  & 0.03 & \\
\hbox{[Al/Fe]}     &  0.08  &  0.00 & 0.07  & 0.07 &  \\
\hbox{[P/Fe]}      &  0.01  &  0.00 & 0.00  & 0.01 &  \\
\hline
\noalign{\smallskip}
\noalign{\smallskip}
\hbox{} & \hbox{} & \hbox{NGC~6569-2M18134025-3149477} & \hbox{} & \hbox{}  &\\
\noalign{\hrule\vskip 0.1cm}
\hline
\noalign{\smallskip}
\hbox{[C/Fe]}  & 0.03 & 0.08 &  0.00 & 0.09 & \\
\hbox{[N/Fe]}  & 0.20 & 0.00 &  0.00 & 0.20 & \\
\hbox{[O/Fe]}  & 0.25 & 0.00 &  0.00 & 0.25 & \\
\hbox{[Na/Fe]} & 0.02 & 0.02 &  0.00  & 0.03 & \\
\hbox{[Al/Fe]} & 0.10 & 0.04 &  0.05  & 0.12 &  \\
\hbox{[P/Fe]}  & 0.17 & 0.05 &  0.00  & 0.18 &  \\

\noalign{\smallskip} 
\hline 
\end{tabular}
\end{table}

\end{appendix}

\end{document}